\documentclass[]{aa}

\usepackage{graphicx}

\newcommand{\kms}{km$\,$s$^{-1}$}
\newcommand{\etal}{et al.}

\newcommand{\Ha}{\hbox{H$\alpha$}}

\newcommand{\CO}{\hbox{$^{12}$CO}\ } 
\newcommand{\isaac}{{\tt ISAAC}}
\newcommand{\vlt}{{\tt VLT}}
\newcommand{\antu}{{\tt ANTU}}
\newcommand{\debca}{{\tt DEBCA}}
\newenvironment{itemize_bullet}{%
	\begin{itemize}}{\end{itemize}}
 
\begin{document}

\thesaurus{
	11
		(11.01.2;  
		 11.11.1;  
		 11.14.1;  
       11.19.1;  
		 11.05.2;  
		 11.19.2   
		 )
}

\title{Dynamics of embedded bars and the connection with AGN
\thanks{Based on observations collected at the VLT telescope of European Southern
 Observatory, Paranal, Chile, ESO N0 64.A-0076, 65.A-0031) }}

\subtitle{I. \isaac/\vlt\ stellar kinematics}

\author{
        E. Emsellem\inst{1}, 
        D. Greusard\inst{2}, 
        F. Combes\inst{3}, 
        D. Friedli\inst{2,6}, 
		  S. Leon\inst{4},
        E. P\'econtal\inst{1}, 
        H. Wozniak\inst{5}}

\offprints{E. Emsellem (email: emsellem@obs.univ-lyon1.fr)}

\institute{
        Centre de Recherche Astronomique de Lyon, 9 av. Charles Andr\'e,
        69561 Saint-Genis Laval Cedex, France \and
        Observatoire de Gen\`eve, CH-1290 Sauverny, Switzerland \and
        DEMIRM, Observatoire de Paris, 61, Avenue de l'Observatoire, 75014 Paris, France \and
		  Institute of Astronomy and Astrophysics, Academia Sinica, P.O. Box 1-87, Nankang, Taipei, Taiwan, R.O.C. \and
        Observatoire de Marseille-Provence, Laboratoire d'Astrophysique de Marseille,
2 Place Le Verrier, F-13248 Marseille Cedex 4, France \and
	Gymnase de Nyon, CH-1260 Nyon, Switzerland
}

\date{Accepted 2000 December 21. Received 2000 October 27}

\maketitle
\markboth{Emsellem et al.: \isaac\ observations of embedded bars. I}{}

\begin{abstract}
We present new stellar kinematical profiles of four galaxy hosts of active galactic nuclei, using
the \CO\ bandhead around 2.3~$\mu$m with the \isaac/\vlt\ spectrograph.
We find that the nuclear bars or discs, embedded in large-scale primary
bars, have all a decoupled kinematics, in the sense that the maximum
of the rotational velocity occurs in the nuclear region.
In three cases (NGC~1097, NGC~1808 and NGC~5728), the velocity dispersion
displays a significant drop at the nucleus, a rarely observed phenomenon.
We also detect kinematical asymmetries ($m=1$ mode) along the nuclear bar 
major-axis of NGC~1808 and NGC~5728, dynamical counterparts 
of corresponding asymmetries in the surface brightness.
We have derived simple dynamical models in an attempt to fit the
kinematics of each galaxy and reconstruct the full velocity field.
For all four targets, the fits are good, and confirm the presence of
the decoupled nuclear components. These models cannot however reproduce the observed
central drop in the dispersion. We suggest that this drop is due to a
transient cold nuclear disc, fuelled by gas inflow along the bar, that
has recently formed new stars.

\end{abstract}

\section{Introduction}
It is now widely accepted that the energy source of active galactic
nuclei (AGN) originates in the accretion of material onto a central
massive black hole. However, the fuelling mechanism of the central
engine remains unclear. The problem is how to transfer mass from the
galaxy into its very central regions (parsec scale). Several
mechanisms can be invoked to induce the potential perturbation that
could initiate gas inflow. Among them, $m=2$ or $m=1$ modes
tiggered via gravitational perturbation by a companion, 
instabilities developing in the galactic disc or both. It
has also been proposed that the so called ``minor mergers'', in which
a gas rich galaxy and a satellite galaxy are involved could play an
important role in triggering activity in Seyfert galaxies (de Robertis
et al.  \cite{Rob98} and reference therein).

An efficient way to drag significant amount of mass in the central
regions would be the presence of a large scale bar in the host galaxy,
which could initiate strong inflows of gas (e.g. Atha\-nas\-soula
\cite{Lia92}; Friedli \& Benz \cite{Fri93}).  However, it has been
shown in recent studies that if stellar formation is marginally
enhanced in barred galaxies, the presence of an AGN is not correlated
with the existence of a bar in its host galaxy (Mulchaey \& Regan
\cite{Mul97}; Ho et al. \cite{Ho97}).
In fact, if the bar indeed initiates gas inflow, its inner Lindblad resonnance (ILR),
when present, stops the inflow and the gas is redistributed in a disc
inside the ILR radius (e.g. Buta \& Combes \cite{Buta96}). Thus a bar is
clearly an efficient way to drag gas in the central regions ($\sim$
100s pc), but another mechanism must take over to allow this gas to
finally fall onto the AGN.

Shlosman et al. (\cite{Shlosman89}) proposed that instabilities such as secondary bars
could develop in the inner disc, starting again the gas inflow. During the past five years,
a large number of bars within bars have been detected and it becomes now
possible to check statistically the impact of these structures on activity of
galaxies. Whether or not there is a higher fraction of secondary embedded bars
observed among Seyfert galaxies is still a matter of debate
(Wozniak et al. \cite{Woz95}; Friedli et al. \cite{Friedli96}; 
Jungwiert et al. \cite{Jun97}; Mulchaey \& Regan \cite{Mul97};
Regan \& Mulchaey \cite{Reg99}; Greusard et al. \cite{Gre00}). 
However, the lack of a clear correlation could be due to the different
timescales in the involved processes (fuelling, AGN phase, bar dissolution).
Still, bars within bars are often associated with bursts of star formation, 
confined within the nuclear bar, or the nuclear ring encircling it.
Anyway, they do lead some significant evolution in the morphology
and dynamics of the central regions of their hosts.

It is therefore important to trace both the dynamics
and the stellar population of embedded bars, and examine
potential links with the central AGN. So far, embedded bars have 
been observed essentially by optical or NIR imaging, whereas lack
of collecting power has prevented any breakthrough from kinematic
studies. Furthermore, if optical spectroscopy can be used to measure
kinematics in dust-free regions, one has to move to less obscured
wavelengths to map galaxy centres which are almost always very dusty (e.g.
Valentijn \cite{Val90}). As shown by Gaffney et al. (\cite{Gaf95}), the \CO\
absorption features at 2.29~$\mu$m, if not widely used, are
a very efficient tool for measuring stellar kinematics in dusty environments.
The age of the stellar populations can also be approached through
the equivalent widths of these absorption \CO\ features (Doyon et al. \cite{Doy94}).
Together with high-resolution NIR photometry, the kinematics
can provide mass-to-light ratios, that also constrain the age
of the populations.

We have thus undertaken a NIR spectroscopic
study of a sample of Seyfert galaxies with and without double bars,
using the spectrograph \isaac\ mounted on the \vlt/\antu. The aim of
this work, the \debca\ (Dynamics of Embedded Bars and the Connection with AGN)
project, is to characterize the kinematics of stars and gas in the
few 100s inner parsecs, and to constrain the age of the
stellar populations. We have so far obtained long-slit \isaac\ 
spectroscopy of four Seyfert galaxies with double bars. 
In the present paper, the first of a series, we present the stellar kinematics
extracted from these data, and discuss it in the light of
simple dynamical models. A detailed discussion regarding the
stellar populations is reported to a forthcoming paper (Greusard et al. \cite{Gre01}, 
in preparation, hereafter Paper~II).

\section{The \debca\ sample}
   \label{sec:Sample}

The main goal of the \debca\ project is to study the link between the
nuclear kinematics of the host galaxy and the fuelled central
engine. We have thus compiled a list of single- and double-barred
galaxies according to a few simple criteria:

\begin{itemize_bullet} 
\item $-70\degr\ < \delta < +20\degr\ $ where $\delta$ is the
declination of the target, for observations with the VLT.
\item $30\degr\ < i < 60\degr\ $, where $i$ is the inclination
angle. This ensures that the bar is clearly visible, but with an
inclination high enough to have significant velocity signatures.
\item $D < 40$~Mpc, which constrains the minimum intrinsic physical
scales ($0\farcs5 \sim 100$~pc) we can probe.
\end{itemize_bullet}
We have then selected a small sample of 12 galaxies,
where both Seyfert~1 and 2 were required. Four of the targets are
non-active galaxies, to be used as a ``control sample''. In
the following Sections, we describe the observations and analysis of
the first four targets of our list, three of them being 
generally classified
as Seyfert~2's, and the fourth being a Seyfert~1 (Table~\ref{tab:optdata}).
However evidence for the presence of a Seyfert nucleus in NGC~1808
is weak, this galaxy being better defined as a starburst galaxy:
indeed optical and near-infrared spectra 
and more recent HST images strongly support the latter
classification (e.g. Kotilainen \etal\ \cite{Kot96}).
NGC1097 and NGC 5728 are both classified
as Seyfert 1's in the Veron-Cetty \& Veron (\cite{Ver93}) catalogue.
Nevertheless, and although NGC~1097 does indeed contain an obscured broad-line 
region, characteristic of a hidden Seyfert~1 nucleus, 
the apparent nuclear activity of NGC~1097 is weak.
This is also the case for NGC~5728 which was mentioned
by Wilson \etal\ (\cite{Wil93}) as a support for the unified model 
of Seyfert galaxies. We are therefore keeping NGC~1097 and
NGC~5728 in the Seyfert 2 class.

\begin{table*}[ht]
\caption[ ]{Optical data for the 4 galaxies.}
\begin{flushleft}
\begin{tabular}{lccccclccrcrr}  \hline
Galaxy & $D$ & $M_B$  & Diameter  & Sey & Bar$^1$& Type & V$^{Leda}_{sys}$ & V$^{obs}_{sys}$ &PA$_{disc}$& $i$ & PA$_{pb}^2$ & PA$_{nb}^3$\\
 &  [Mpc] & [mag]   &  [$\,\arcmin \times \arcmin\,$]  &      &       &  &  [\kms]    &  [\kms] & [$\circ$] & [$\circ$] & [$\circ$] & [$\circ$]\\
\hline
NGC\,1097 & 16.8 &10.2&9.3 $\times$6.3&S2&DB & SBb & 1273 & 1240 & 130&37& 138 & 30\\
NGC\,1365 & 18.6 &10.3&11.2$\times$6.2&S1&DB & SBb & 1653 & 1628 & 32 &57& 91 & 45\\
NGC\,1808 & 10.9 &10.8&6.5 $\times$3.9&SB$^{\star}$&DB & SABb& 1003 & 1015 & 133&70& 143 & 157\\
NGC\,5728 & 37.0   &12.4&3.1 $\times$1.8&S2&DB & SABa& 2789 & 2836 & 30 &55& 33 & 79\\
\hline
\end{tabular}
\end{flushleft}
$^1$ DB means Double Bar\\
$^2$ PA of the Primary Bar\\
$^3$ PA of the Nuclear Bar\\
$^{\star}$ Starburst galaxy: evidence for a Seyfert nucleus in NGC~1808 is weak, see text.\\
\label{tab:optdata}
\end{table*}

\section{The data}
   \label{sec:Obs}

        \subsection{Observations}
We obtained 12 hours of observing time in service mode (Period~63)
and 2 nights in visitor mode (Period~64) at the \vlt/\antu\ with the
\isaac\ spectrograph. We have used the long-slit mode of \isaac\
in its Short Wavelength configuration: characteristics are
given in Table~\ref{tab:isaac}. Our spectral domain includes
the first three or four (depending on the redshift of the galaxy)
\CO\ bandheads, around 2.3~$\mu$m.

We defined Observing Blocks ({\tt OB}) of 14 ({\tt NDIT}) exposures of 180s
({\tt DIT}) each, operated in noding mode: the objects in exposures ``$A$'' and ``$B$'' were
centred on the first and last third of the NIR array respectively.
This (classical) procedure allowed us to have an excellent sky subtraction,
using $A-B$ and $B-A$ differential exposures as working frames for the
reduction. 

We have observed 4 galaxies in our sample of 12, namely
NGC~1097, NGC~1365, NGC~1808 and NGC~5728.
Details for each galaxy are given in Table~\ref{tab:data}.
Each galaxy was originally supposed to be observed for a total
of 10080s (2.8 hours = 4 {\tt OB}s). However, as mentioned
in Table~\ref{tab:data}, we had to discard a number of exposures
due to technical problems mainly due to:
\begin{itemize_bullet}
\item Spatial drifts of the slit during the {\tt OB}.
\item Slit jumps which resulted in a tilted slit both
with respect to the NIR array and the noding. 
\end{itemize_bullet}

We also observed a set of stellar kinematical templates (typically G, K and M giants)
to be used for the kinematical measurements, and solar
type stars for the correction of telluric features (see Maiolino et al.
 \cite{Mai96}). 

In our spectral domain, there are no OH lines, 
generally useful to perform a wavelength calibration of the exposures.
We had thus to rely on independent arc lamp exposures to perform our
wavelength calibration. In this context, we asked individual arc exposures
{\em during the night}.     
\begin{table}
  \caption{Instrumental setup of \isaac}
  \label{tab:isaac}
  \begin{flushleft}
  \begin{tabular}{ll}
  \hline
  \multicolumn{2}{c}{\isaac\ SW mode} \\
  \hline
  Slit & $0\farcs6$$\times$120\arcsec \\
  Spatial sampling & $0\farcs147$ \\
  Spectral sampling & 1.19 \AA \\ 
  Spectral resolution & 4478 \\
  Spectral FWHM & 67 \kms \\
  Wavelength interval &  1200 \AA\ centred at 2.336~$\mu$m\\
  \hline
  \end{tabular}
  \end{flushleft}
\end{table}
\begin{table}
\caption[]{DEBCA-\isaac\ data characteristics. The observation period is given in column ``P''.
 Numbers of OBs are indicated as
used / discarded respectively.  ``Exp'' is the total exposure time on target.
FWHM$_{\star}$ corresponds to the mean seeing.
The orientation of each slit with respect to the nuclear bar major-axis is given 
in column~3, the PA of the slit is given in Column 4.
}
        \begin{center}
        \begin{tabular}{|lccr|ccc|}
        \hline
        Galaxy & P & axis & PA      & \# OB & Exp.  & FWHM$_{\star}$ \\
               &   &      & [$\degr$]&       & [min] & [$\arcsec$]\\
        \hline
        \hline
        NGC~1097 & 63 & //  & 29.5 & 4\ /\ 0\  & 168 & 0.8 \\
         & & $\perp$ & 119.5 & 5\ /\ 0\  & 168 & 0.7 \\
        NGC 5728 & 63 & //  & 264.5 & 4\ /\ 0\ & 168 & 0.6 \\
         & & $\perp$ & 354.5 & 4\ /\ 0\  & 162 & 1.5 \\
         \hline
         \hline
        NGC 1365 & 64 & // & 45.5 & 4\ /\ 0\  & 162 & 1.0 \\
         & & $\perp$ & 135.4 & 4\ /\ 1\  & 162 & 1.0 \\
        NGC 1808 & 64 & //  & 335.5 & 5\ /\ 0\  & 210 & 0.7 \\
         & & $\perp$ & 65.5 & 3\ /\ 2\  & 112 & 0.6 \\
         \hline
        \end{tabular}
        \end{center}
        \label{tab:data}
\end{table}

\subsection{Data reduction}

In the following paragraphs, we give a brief description
of the reduction and analysis procedure we applied to our data.
We emphasize some of the problems we encountered on the way,
most of them linked with instrumental issues (all data
were taken prior to the major overhaul in Feb. 2000). All the reduction
processes were applied using the {\tt IRAF} and {\tt MIDAS} packages,
as well as a few low level routines from the {\tt Eclipse} package.

Since we observed in nodding mode, we
used the differential comparison ($A-B$ and $B-A$) to subtract
the dark, bias and sky contribution from all exposures.
The data were then flat-fielded using a previously prepared
master flat field image: variations of up to 5\% were measured
on the flat fields during a night. We then corrected for the distortion 
along and perpendicular to the slit, using the star-trace exposures provided
by ESO, and associated arc lamps. Systematic residual (low frequency) distortion
were of the order of 0.2-0.3 pixel, not fully satisfactory, 
but sufficient in the context of our program. It seems that these residuals cannot
be further damped, as the distortion pattern varied on a medium time range
at the time of the observations (which means that the 
star-trace exposures were not stable enough).

The data were then wavelength calibrated.
As already mentioned, there are no sufficiently bright OH lines 
in our spectral domain to allow any spectral calibration, and we had
to rely on independent arc lamp exposures. Unfortunately, at the 
time of the observations, there was a (known) problem with the 
dispersor which seems to shift from one {\tt OB} to the next,
following an automatic software initialisation. We have indeed 
observed some significant shifts (typically a few 
tenths of a pixel) along the dispersion direction 
between successive {\tt OB}s. This is critical for our program as 
we are looking for a velocity accuracy of $ < 5$~\kms, 
a third of a pixel. This problem was solved by using sky emission lines 
to correct for any residual zeroth order shift.

\begin{figure}
\resizebox{8cm}{!}{\includegraphics{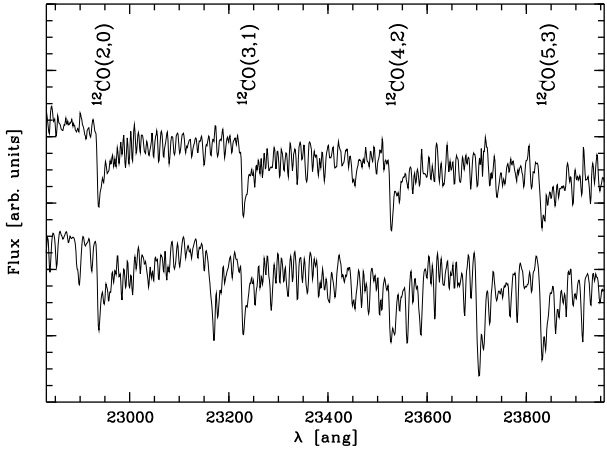}}
 \caption{\isaac\ (aperture) spectrum of HD~16492, a K0 giant,
	 before (bottom) and after (top) correction for the telluric absorption. The main
 \CO\ lines are identified.}
 \label{fig:telluric}
\end{figure}

Individual exposures are then combined, after careful recentring,
and corrected for telluric absorption using a solar type stellar template as described
in Maiolino et al. (\cite{Mai96}), and taking into account the difference
in line depth (depending on e.g. the differential airmass). The result is illustrated
in Fig.~\ref{fig:telluric} for a K0{\sc III} star. 
The present data reduction only provided us with a relative flux calibration,
sufficient for kinematical purposes.

\subsection{Kinematical analysis}
\label{sec:kin}

The (stellar template and galaxy) spectra were finally rebinned 
in $\ln{(\lambda)}$ to be sampled with constant bins in velocity 
space.  We first binned spectrally by a factor of 2
as this leads to a pixel of about 31~\kms, properly sampling
the original spectral resolution of the data (see Table~\ref{tab:isaac}).
We also binned the data spatially along the slit to ensure
a minimum signal to noise ratio of 20, required to extract the
stellar kinematics.
We then performed a continuum subtraction using a low order polynomial.
A refined version of the Fourier Correlation Quotient (Bender 1990)
was used to derive the line-of-sight velocity distribution and to measure
the first two velocity moments ($V$ and $\sigma$): we used
different templates and checked that the resulting kinematics
were not significantly affected by template mismatching.
Measurements of higher order Gauss-Hermite moments will wait for the
building of optimal templates (Paper~II).
The central velocity value was assumed to be the systemic velocity
and subtracted from each individual velocity profile\footnote{Systemic velocities
derived from the minor and major axis of a galaxy
were always found to be consistent with each other, within the error bars.}.

We derived formal errors for the kinematics using a Monte Carlo
approach. Fixing the signal to noise ratio and the velocity
dispersion, we made 500 realisations of simulated broadened spectra,
measured the kinematics via FCQ, and derived the resulting
standard deviation for $V$ and $\sigma$, $S_V$ and $S_{\sigma}$
respectively. $S_V$ and $S_{\sigma}$ were tabulated for 5 values 
of $\sigma$ and 40 values of the signal to noise. We then
derived the errors for individual data points via interpolation.

\section{Kinematical results}
   \label{sec:Res}
 
The kinematical profiles obtained for the four galaxies are
displayed in Figs.~\ref{fig:n1097}, \ref{fig:n1365}, \ref{fig:n1808} and \ref{fig:n5728}.         
In each figure, the lengths of the sketched slits overimposed on the image
correspond to the presented profiles.  

For the interpretation of the velocity profiles, it is important
to keep in mind that the spectra have been taken along
slits parallel and perpendicular to the nuclear bars 
(respectively named Slit\,1 and Slit\,2 hereafter).
The major and minor axis of the galaxies do not coincide 
with that of the bars, so that we do expect some
rotation along both axis, due to the inclination effect.
The various position angles and inclinations are displayed
in Table~\ref{tab:optdata}.

	\subsection{Individual descriptions}
\label{sec:co}
{\em NGC~1097}, Seyfert 2, $1\arcsec \sim 81$pc:
\begin{figure}
\resizebox{8cm}{!}{\includegraphics{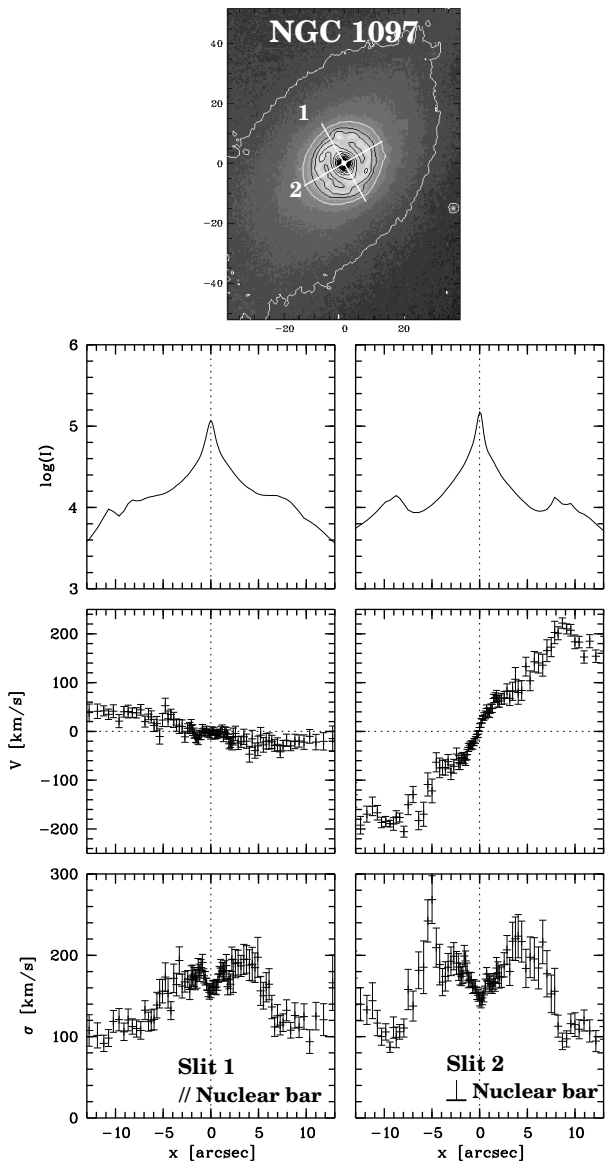}}
\caption{ Kinematical profiles of NGC 1097. 
From top to bottom: NIR image of the galaxy centre (North is up, East is left), 
indicating the 
positions and lengths of the two slits, parallel and perpendicular to the nuclear bar
(the labels 1 and 2 overimposed on the image indicate positive abscissa);
Luminosity profiles (in log) along the two slits; 
Velocity and dispersion profiles with error bars representing
$3\cdot S_V$ and $3\cdot S_{\sigma}$ values (see Sect.~\ref{sec:kin}).}
\label{fig:n1097}
\end{figure}

The luminosity profile along Slit\,1 falls down to a plateau near the
end of the nuclear bar, and then decreases longward $\sim 10\arcsec $
with a roughly exponential law characteristic of a disc population.
This is fully consistent with the $10.3\arcsec$ semi-major axis extent
of the nuclear bar mesured by Friedli et al. (\cite{Friedli96}). The
luminosity along Slit\,2 follows the same behaviour, but with a bump
just after the plateau due to its well--known circumnuclear 
clumpy ring (actually a tightly wound spiral structure in the NIR; 
e.g. Kotilainen et al. \cite{Kot00}).

The velocity profile along Slit\,1 is quite flat, reflecting the fact that
it is nearly perpendicular to the kinematical line of nodes 
(which we assume to be given by the major-axis
photometric position angle of the outer disc of the galaxy). The global shape of the
rotation curve along Slit\,2 roughly resembles the \Ha\ velocity curve
derived (along a position angle of $130\degr$) by Storchi-Bergmann et
al. (\cite{Stor96}). The maximum stellar velocity along Slit\,2 
($V_{\rm max,2}$ $\sim 210$ \kms) is reached in the
circumnuclear ring, similarly to the ionised gas for which Storchi-Bergmann et
al. (\cite{Stor96}) measures maxima of $V_{\rm max}$[\Ha] $\sim 225$ \kms.
We thus measure a roughly constant stellar velocity gradient of $\sim$
290~\kms\,kpc$^{-1}$. Our good spatial resolution however allows to reveal a richer
velocity structure. Inside $R = 5 \arcsec$, the velocity profile along
Slit\,2 exhibits an S-shape with nearly flat ends. 
Those plateaus in the velocity correspond to maxima in the
dispersion profile ($\sigma_{2} \sim 220$ \kms), whereas the inner part is
characterised by a quite surprising drop in the dispersion (down to
$\sigma_{2} \sim 145$ \kms at the centre). 
Velocities increase almost linearly from a radius of $5\arcsec$
reaching a maximum near the edge of the circumnuclear ring at $\sim 9\arcsec$,
where they then starts to decrease. Outside $5\arcsec$, the dispersion decreases 
outwards down to $\sim 95$~\kms. Note that the dispersion 
drop and local maxima in the dispersion are also present along Slit\,1. 

\medskip
{\noindent \em NGC~1365}, Seyfert 1, $1\arcsec \sim 90$pc:

The Seyfert~1 nucleus of NGC~1365 dominates the light
in the central arcsecond, and thus strongly dilutes the absorption \CO\ 
bandhead: this prevented us to derive any meaningful kinematics in
this region. We will deal here only with the profiles outward $R \geq 2 \arcsec$.

Like in NGC~1097, the flatness of the velocity profile along Slit~2 is
a consequence of the slit orientation with respect to the line of
nodes. The central kpc morphology of this galaxy is disturbed by an
intense star formation (see Lindblad
\cite{Lind99} for a review on this object). It is thus
difficult to see the signature of the bar in the luminosity
profile. Ellipse fitting on $H$-band isophotes provided by
Jungwiert et al. (\cite{Jun97}) gave a rough estimate of the extent
of the presumed secondary bar: $\sim 9\!-\!10 \arcsec$. 
However, high resolution near-infrared images
of the central region of NGC~1365 recently obtained with NICMOS/HST, and 
the VLT (ISAAC and FORS1) suggest that the ellipticity of the component
detected in the central 10\arcsec\ is solely due to the inclination
of the galaxy (the photometric major-axis being thus coincident with
the line of nodes). There are therefore no evidence left for the
presence of a nuclear bar. We then simply interpret the observed 
flattened system in the centre as a nuclear disc, well circumvented by a ring-like 
(and spiral arm) structure at a radius of $\sim 7\arcsec$. Inside 
this radius, the velocity increases up to its
maximum value ($V_{\rm max,1} \sim 175 $~\kms at $R \sim 5 \arcsec$)
with a steep gradient ($\sim 390$~\kms\,kpc$^{-1}$), and then remains
roughly constant until the end of the disc. The dispersion along both axis 
shows no clear structure: it remains nearly constant inside the nuclear disc
with a mean $\sigma_{1} \sim 100$~\kms. There may be a slight increase
outwards up to $\sigma_{2} \sim 120\!-\!130$~\kms, but this is within the
error bars.
\begin{figure}
\resizebox{8cm}{!}{\includegraphics{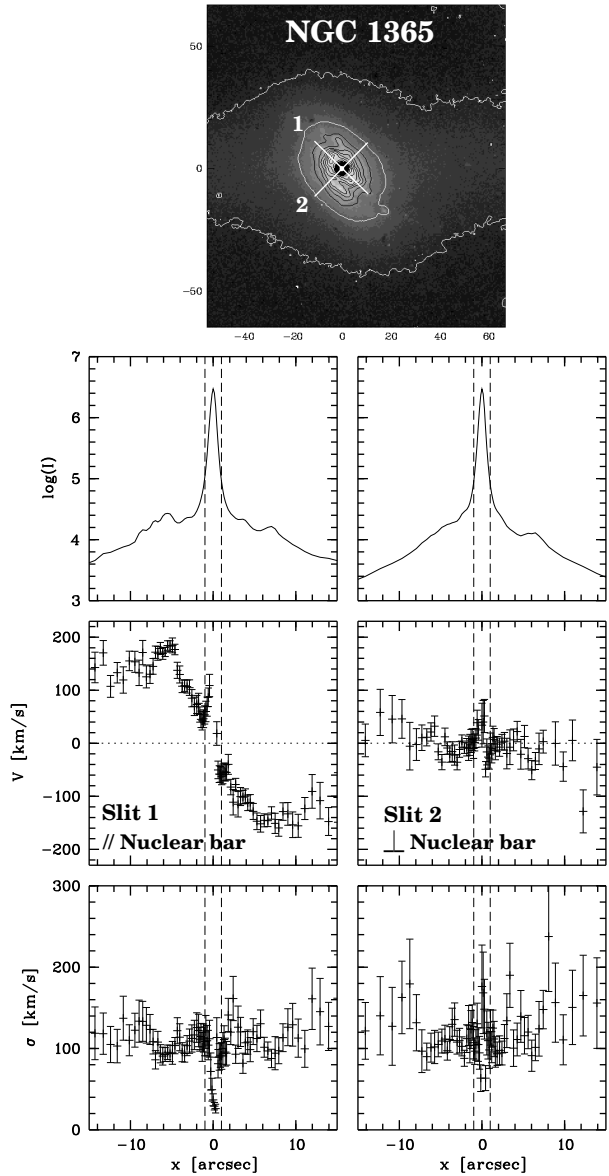}}
 \caption{ Same as Fig.~\ref{fig:n1097} for NGC 1365. The light in
 the central arcsecond (region marked by the vertical dashed lines)
is completely dominated by the non-thermal contribution
 of the Seyfert~1 nucleus, thus preventing us to derive any meaningful
 kinematics in this region.} 
 \label{fig:n1365}
\end{figure}

\begin{figure}
\resizebox{8cm}{!}{\includegraphics{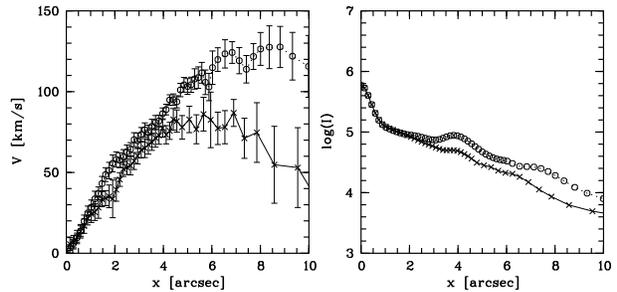}}
 \caption{Velocity (left, absolute values) and surface brightness (right) profiles along Slit\,1 of
NGC~1808: the crosses and solid lines correspond to the north--west side,
 the circles and dotted lines to the south-east side.} 
 \label{fig:m1_n1808}
\end{figure}

\medskip
{\noindent \em NGC~1808}, Seyfert 2, $1\arcsec \sim 53$pc:

The central kpc of this galaxy, disturbed by 'hot spots' of star
formation (e.g. Kotilainen et al. \cite{Kot96}), is the brightest of
our sample in the $K$-band, hence providing the nicest kinematic profiles.
Again, we estimate the nuclear bar length to be $\sim 6
\arcsec$ from Jungwiert et al. (\cite{Jun97}), with an axis ratio
around 0.5. The velocity profiles along both axis 
show an increase up to the end of the nuclear bar and then a decrease. 
Slit\,1 velocity profile is significantly asymmetric with respect
to the systemic velocity outside $2\arcsec$, 
with $V_{min,1} = -124$~\kms and $V_{max,1} = 81$~\kms: 
this asymmetry is also clearly present
in the surface brighness profile (Fig.~\ref{fig:m1_n1808}) but
does not appear in the dispersion curve.
\begin{figure}
\resizebox{8cm}{!}{\includegraphics{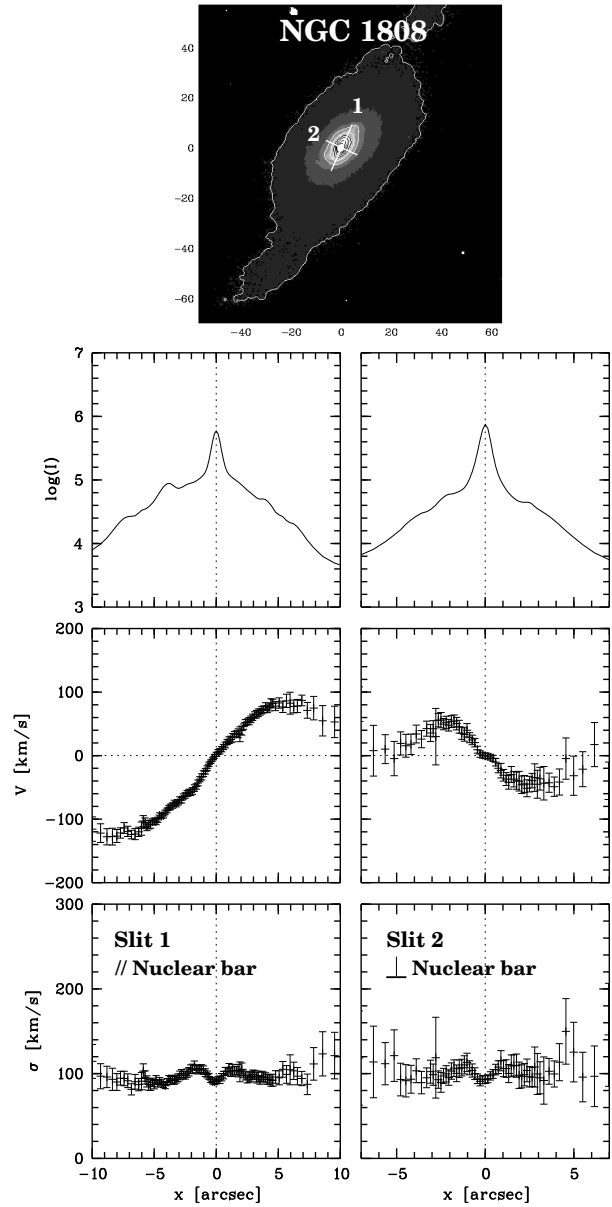}}
 \caption{ Same as Fig.~\ref{fig:n1097} for NGC 1808. } 
 \label{fig:n1808}
\end{figure}
\begin{figure}
\resizebox{8cm}{!}{\includegraphics{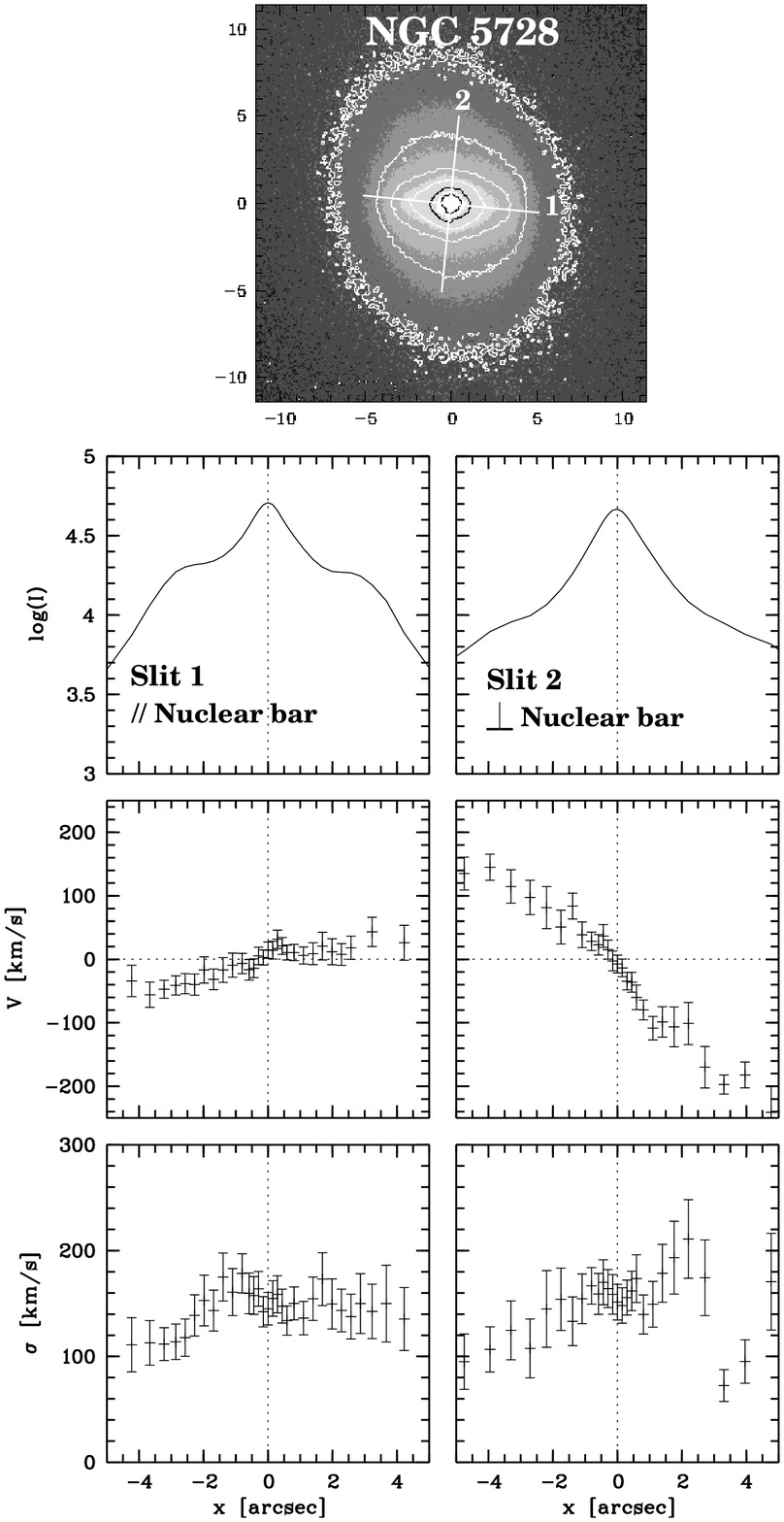}}
 \caption{ Same as Fig.~\ref{fig:n1097} for NGC 5728. Note the different spatial
 extent of the major and minor axis plots.} 
 \label{fig:n5728}
\end{figure}
Along Slit\,2, the velocity profile is roughly symmetric with local extrema
V$_{\rm min,2} \sim -45 $~\kms and V$_{\rm max,2} \sim 53 $~\kms 
at radii of $\pm 2\arcsec$, followed by a decrease outwards down to systemic
velocity. The mean slopes in the rising parts are
$\sim 300$~\kms\,kpc$^{-1}$ and 260~\kms\,kpc$^{-1}$ along Slit\,1 and Slit\,2 respectively. 
There is however a distinct kink in the velocity profile along Slit~1
at a radius of $-2\arcsec$, with no symmetric counterpart.
The velocity gradient changes inside $0\farcs4$ of Slit~2,
to only $\sim 140$~\kms\,kpc$^{-1}$, a real feature considering 
the final spatial resolution for this slit 
($\sim 0\farcs6$ FWHM; see Table~\ref{tab:data}).
The dispersion profiles remain in the range 80--120 \kms and
exhibit a trend similar to the one of NGC~1097: an increase toward the
centre followed by a significant drop this time inside 1\arcsec. The magnitude of
this drop is low ($\sim 15 $ \kms) but real.

\medskip
{\noindent \em NGC~5728}, Seyfert 2, $1\arcsec \sim 179$pc:
NGC 5728 is the faintest galaxy in our sample, however the signal-to-noise
is enough to have a good measure of the velocity profiles in the central 5\arcsec.
Once again, there are clear signatures of a decoupled dynamical component in the
central kinematics. The maximum velocity gradient is observed along
Slit\,2 as expected from the position angle of the line of nodes 
(see Table~\ref{tab:optdata}), similarly to the case of NGC~1097.
Slit\,1 for NGC~5728 is very close to the slit used 
by Prada \& Gutierrez (\cite{Pra99}, hereafter PG99; PA$=86\degr$), 
our velocity profile being consistent with theirs.
The velocity and velocity dispersion profiles are slightly asymmetric
along both slits. The $K$ band surface brightness profile 
also exhibits an asymmetry along Slit\,1 at the edge of the nuclear bar.
There is a dip in the dispersion profiles of NGC~5728,
with a central value of 147~\kms, although it is less convincing than in the cases
of NGC~1097 and NGC~1808. This value is slightly smaller than but within the error bar of
the one derived by PG99. We do not detect any double component in our LOSVDs,
in apparent contradiction with the data of PG99. We observe
a high excitation [Ca{\sc VIII}] emission line, burried within the second \CO\ absorption
feature. The emitting region is restricted to the central spectra, and is consistent
with an unresolved point-like source, thus certainly linked with the AGN.
This point will be examined in details in a forthcoming paper (Paper~II).
      
\subsection{Global results}

The first striking result of those observations is that, in all 4 observed targets,
the rotational velocity reveals a maximum inside the nuclear bar (or disc for NGC~1365)
and then decreases,\footnote{In the case of NGC~5728, we hardly reach the 
end of the nuclear bar so that the velocity decrease is less 
obvious, but see PG99.} showing that the nuclear region is a well decoupled dynamical
component of the galaxies. For the three cases with nuclear bars
(NGC~1097, NGC~1808 and NGC~5728), this follows suggestions made from
photometric studies as no preferential angle was observed 
between the two bars (Greusard et al. \cite{Gre00} and references therein).
The existence of such structures could be doubted when
dealing with a galaxy like NGC~1808, where there are numerous
clumps of star forming systems within the central arcseconds.
But even in that case, the NIR photometric elongation 
embedded within the ring present in the WFPC2/HST (archival) images strongly
suggests the presence of a nuclear bar. This point is further examined
in the light of dynamical models (Sect.~\ref{sec:Model}).
This is therefore the first direct confirmation of the dynamically 
decoupled nature of nuclear bars.

The second surprising result comes from the dispersion profiles: they
exhibit a significant drop at the centre (but again we cannot say
anything concerning NGC~1365).  This is particularly clear in the
cases of NGC~1097 and NGC~1808. We have checked that the dilution of
the lines by any featureless continuum component does not affect the
dispersion (and velocity) as long as the \CO\ lines remain strong
enough. We do indeed see some dilution and changes in the \CO\ line
strength, but this does not affect our result.
We have also checked that the observed central dispersion drop 
is not due to a template mismatching effect (Paper~II).

\section{Modelisation of the kinematics}
   \label{sec:Model}
For each galaxy, we wish to reconstruct the entire velocity field, projected on the
sky, and constrain it with the observed velocity profiles
along the short and long axis of the nuclear bar (or disc for NGC~1365).
Three of the four galaxies observed here have embedded nuclear bars,
and the orbits are then not expected to be circular.
To interpret these profiles, in a first approximation, we build simple 
models of the orbital structure, based on the epicyclic theory,
assuming that the departures from circular motions are small. 

\subsection{Linear approximation}
We introduce the usual coordinate system ($\xi$, $\eta$), rotating
with the angular speed of rotation $\Omega - \Omega_p$ in the
frame co-rotating with the bar perturbation ($\Omega_p$).
$$
r = r_0 + \xi 
$$
$$
\theta = \theta_0 +  (\Omega - \Omega_p) t + \eta/r_0
$$
where $\theta$ is the azimuthal angle in the rotating frame.

The bar potential is modelled by the function:
$$
\Phi(r,\theta) = -\Phi_2(r) \cos{ 2\theta} + ....
$$
with small harmonic terms ($\Phi_4, \Phi_6 \le \Phi_2$). 
The shape of the bar potential is taken from the $k=3$ bar of the
surface density-potential pairs generated by Kalnajs (\cite{Kal76}):
$$
\Phi_2(r) = 0.53 q_{bar} x (1.-2.5 x+2.1875 x^2-0.65625 x^3)
$$
when $x = r^2/r_{bar}^2 < 1 $. Outside $r_{bar}$, the potential
is extrapolated by continuity, with an inverse power law (in $r^{-4}$). 
The strengths of the bars $q_{bar}$ are chosen such that the perturbations
in the velocity field match at best the data. The strength of the perturbation
is best quantified by the maximum over the disc of the ratio of
the tangential force to the radial force, P2$_{max}$. This quantity is
given in Table~\ref{tab:modpar} for the four galaxies.

The equations of motion are linearized; in order to take into account
the transition at the inner Lindblad resonances, an artificial frictional 
force is introduced, with a damping coefficient $\lambda$, as is
usually done to simulate gas orbits (Lindblad \& Lindblad \cite{Lin94}, Wada \cite{Wada94}).
The motion is that of an harmonic oscillator, forced by
an imposed external perturbation. 
The equations can be solved, at the neighborhood of the ILR
(and OLR) and far from corotation, and give the coordinates
and velocities of the orbit of the guiding centre, the epicyclic 
motions around this centre being damped by the frictional force.

This formulation (see Appendix~A) accounts for the change of
orientation of orbits at the crossing of resonances (parallel or
perpendicular to the bar), and when the damping parameter 
 $\lambda$ is not zero, of a gradual orientation change, corresponding
to the gas spiral arms. We also use models with spiral configurations,
since some young supergiant stars, just formed out of the gas,
share its dynamics (and are indeed observed in the NIR range).
  
\subsection{Galaxy models}

The mass model for the spiral galaxies 
is made of three components: 
\begin{itemize}
\item a small bulge, of Plummer shape potential
$$
\Phi_b(r) = - {{G M_b }\over {\sqrt{r^2 +r_b^2}}}
$$
with characteristic mass $M_b$ and radius $r_b$;
\item a Toomre disc, representing the main
stellar disc, of surface density
$$
\Sigma(r) = \Sigma_0 ( 1 +r^2/r_d^2 )^{-3/2}
$$
truncated at r$_{t}$;
\item a nuclear disc,
 which corresponds to the
decoupled nuclear bar, and is required to account for the
large gradients of velocity profiles, observed at kpc scale
in the present work. Its shape is also taken as a Toomre
function:
$$
\Sigma(r) = \Sigma_0 ( 1 +r^2/r_{nd}^2 )^{-3/2}
$$
with a characteristic mass $M_{nd}$ and radius $r_{nd}$.
\end{itemize}
Let us emphasize that within the centre of spiral galaxies,
the presence of dark matter is not required
(e.g. Freeman \cite{Free92}): the
spherical dark matter halo, usually added to flatten
rotation curves at large radii, have such a large characteristic
radius that it has no influence on the inner parts considered here.
All parameters are displayed in Table \ref{tab:modpar}. 
From these analytical components, it is easy to compute the
critical velocity dispersion for axisymmetric instability at 
each radius. We assume that the radial velocity dispersion $\sigma_r$
is everywhere proportional to this critical velocity, with a constant
Toomre parameter $Q$ as a function of radius. The value of
$Q$ is also indicated in Table \ref{tab:modpar}. The tangential
velocity dispersion  $\sigma_{\theta}$ is assumed to verify
the epicyclic relation:
$$
\sigma_{\theta} = \frac{\kappa}{2\Omega} \sigma_r.
$$ 
The disc plane is assumed to have a constant scale-height $h_z$
with radius, and the z-velocity dispersion is derived from the
surface density in the disc $\Sigma(r)$ by:
$$
 \sigma_z^2(r)=2 \pi G \Sigma(r) h_z.
$$

For each observed slit, we have computed the velocity
dispersion along the line-of-sight in combining the local dispersion,
and the contribution of the velocity gradient in the observed 
spatial resolution. Even in the very centre of the galaxies, this 
second contribution was always negligible.

\begin{figure}
\resizebox{8cm}{!}{\includegraphics{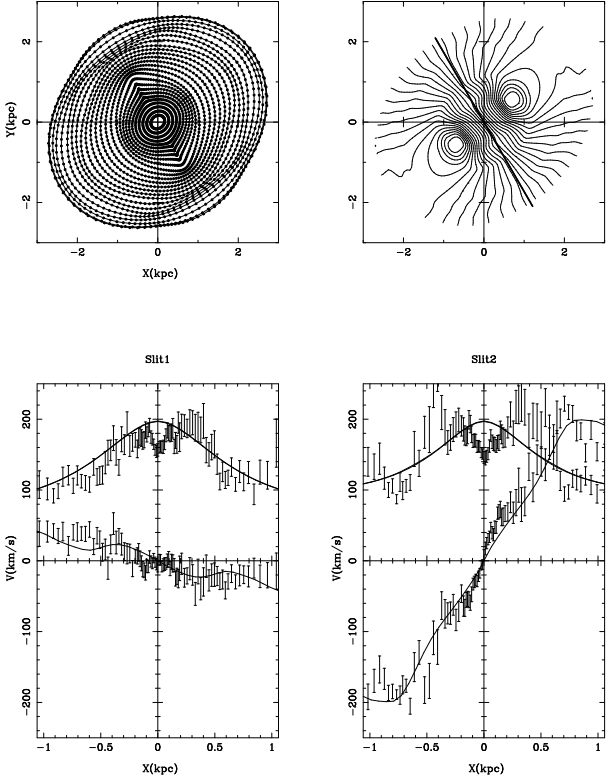}}
 \caption{ Model velocity profiles for NGC 1097.
{Top left:} the shape of the orbits in the linear epicyclic approximation,
projected on the sky plane.
{Top right:} the deduced velocity field, with the orientation of the nuclear
bar indicated.
 {Bottom:} The corresponding velocity profiles, along the nuclear bar
(left), and perpendicular to it (right) overimposed on the
\isaac\ kinematical profiles and their corresponding error bars. 
Two lines are plotted for the modeled velocity
dispersion profiles, including or not the velocity gradients in the 
resolution elements. Most of the time, the two profiles are coinciding.}
 \label{fig:mod_1097}
\end{figure}
\begin{figure}
\resizebox{8cm}{!}{\includegraphics{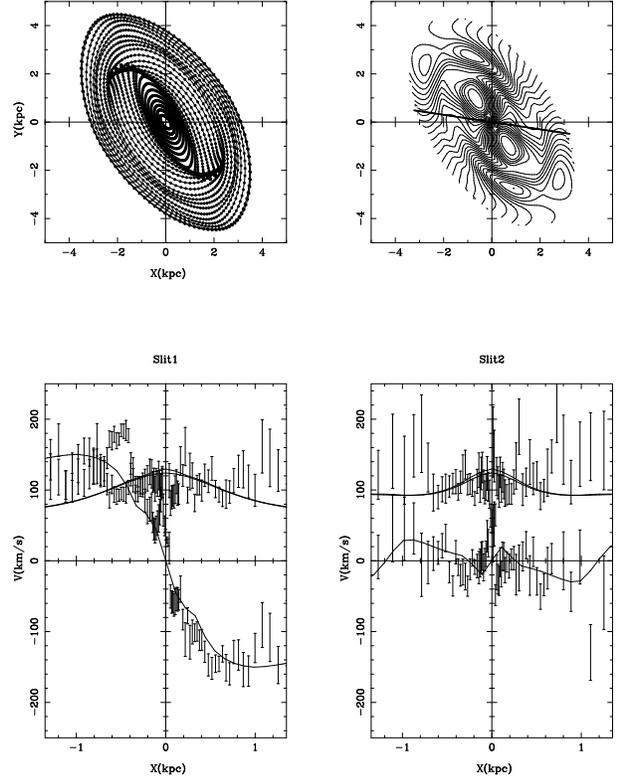}}
 \caption{ Model velocity profiles for NGC 1365.
(see Fig.~\ref{fig:mod_1097} for caption)}
 \label{fig:mod_1365}
\end{figure}
\begin{figure}
\resizebox{8cm}{!}{\includegraphics{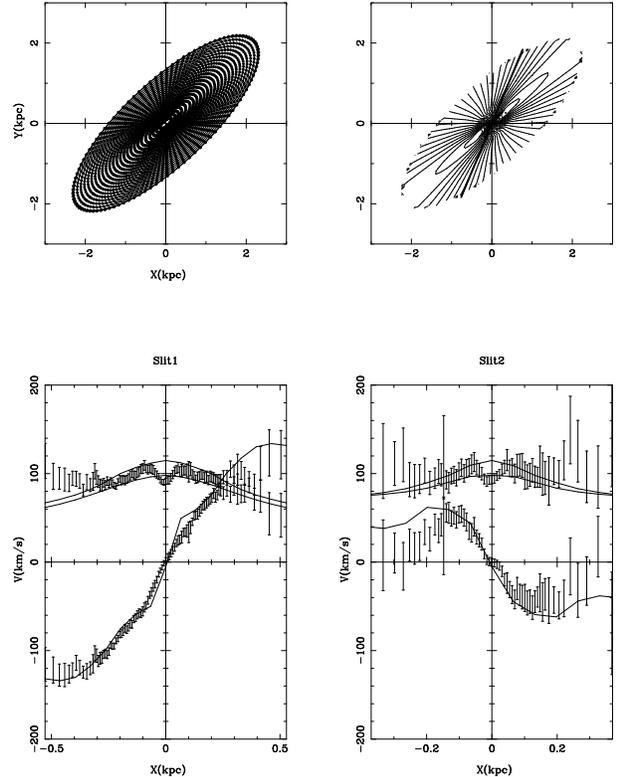}}
 \caption{ Model velocity profiles for NGC 1808.
(see Fig.~\ref{fig:mod_1097} for caption)}
 \label{fig:mod_1808}
\end{figure}
\begin{figure}
\resizebox{8cm}{!}{\includegraphics{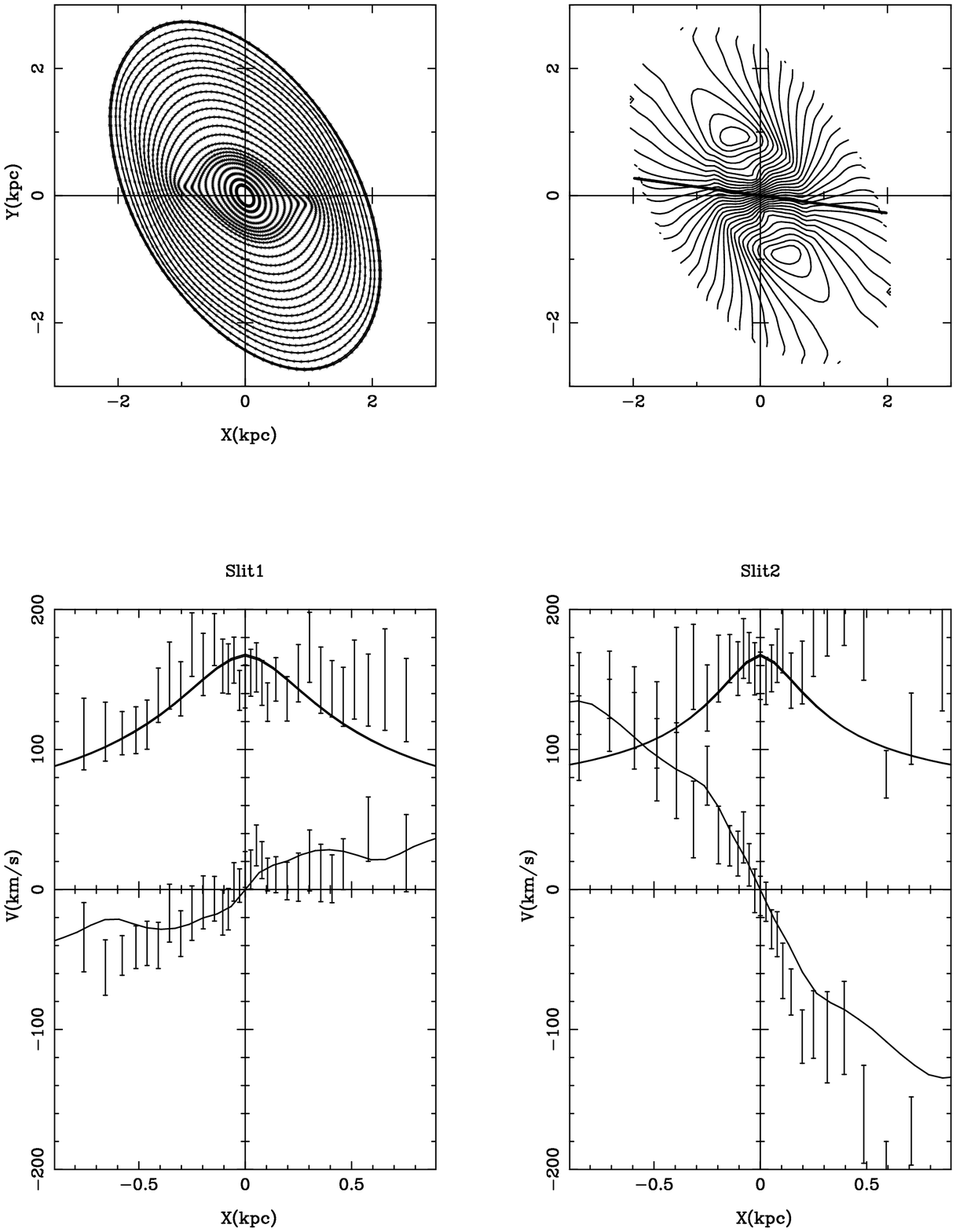}}
 \caption{ Model velocity profiles for NGC 5728.
(see Fig.~\ref{fig:mod_1097} for caption)}
 \label{fig:mod_5728}
\end{figure}

\subsection{Fits of the observations}

The results of the modelisation can be seen in
Figs.~\ref{fig:mod_1097}, \ref{fig:mod_1365}, \ref{fig:mod_1808} and
\ref{fig:mod_5728} for the four galaxies. In all cases, the presence
of the nuclear disc is necessary to account for the large velocities
at the kpc scale. The small bulges allowed by the photometry are
insufficient, and in general bring a negligible contribution to the
rotation curve. To limit the number of free parameters, we have fixed
the mass and radius of the bulges to the statistical relation found
by de Jong (\cite{Jong96}) through NIR photometry. According to the types
of the present galaxies, the bulge-to-disc ratio is 0.1, and the
bulge radius is about 10 times lower than the disc radius.
The main stellar disc are truncated at $r_{disc}$ = R$_{25}$, and
their radial scale-lengths are deduced, assuming a central surface
brightness of 21~B-mag.arcsec$^{-2}$. The scale-heights of the discs
are chosen to be 0.2 times the radial scale-length (e.g. Bottema \cite{Bott93}).
The remaining free parameters to fit are therefore:
\begin{itemize}
\item M$_d$ the disc mass, which is constrained by the outer parts of the 
velocity curve;
\item M$_{nd}$ and $r_{nd}$ the mass and characteristic size of the circumnuclear
disc. These parameters are constrained by the inner parts of the
rotation curve;
\item The strength and radius of the bar, $q_{bar}$ (or equivalently
		P2$_\mathrm{max}$) and $r_{bar}$,
together with the pattern speed $\Omega_p$,
constrained by the shape and amplitude of the velocity profile;
\item $Q$ the Toomre parameter, constrained by the observed dispersion.
\end{itemize}
Since we are concerned here only with the circumnuclear regions, there is
no corotation inside the model, but most of the time the best fit
of $\Omega_p$ is such that there are one or two inner Lindblad resonances.
\begin{table*}[ht]
\caption[ ]{Model parameters for the 4 galaxies}
\begin{flushleft}
\begin{tabular}{lllccclclcccc}  \hline
Galaxy & M$_b$ & $r_b$ & M$_{nd}$ & $r_{nd}$ & M$_d$ & $r_d$ & r$_{t}$ & $Q$ & $h_z$ & $r_{bar}$ &
$\Omega_p$ & P2$_{max}$\\
&10$^{10}$ M$_\odot$& kpc& 10$^{10}$ M$_\odot$& kpc&
 10$^{10}$ M$_\odot$& kpc & kpc&& kpc& kpc&km/s/kpc & \\
\hline
NGC\,1097  & 1.3  &0.6& 1.6&0.9& 13 &5.9 &22 & 1.5 & 1.2 & 3 & 78 & 0.06 \\
NGC\,1365  & 1.0  &0.8& 1.1&1.2& 10 &8.0 &30 & 1.8 & 1.6 & 5 & 25 & 0.23 \\
NGC\,1808  & 0.3  &0.3& 0.9&0.4& 2.7 &2.7 &10 & 1.4 & 0.5 & 0.9 & 60& 0.07 \\
NGC\,5728  & 0.4  &0.4& 0.9&0.6& 3.8 &4.3 &16 & 1.6 & 0.9 & 3 & 50& 0.06 \\
\hline
\end{tabular}
\end{flushleft}
\label{tab:modpar}
\end{table*}

For all galaxies except NGC 1365, the best fit is obtained with
a nuclear bar, oriented differently than the primary bar, and
parallel to the apparent nuclear bar. For NGC 1365 however,
it was better to keep the primary bar potential orientation, 
with its low pattern speed, and rely on the different phase
orientation of the orbits, to form a spiral nuclear
structure, between the two ILRs. This confirms the observation
that NGC~1365 does not include a secondary
bar, as previously claimed, but a decoupled nuclear disc
surrounded by spiral arms within the ILR of the primary bar.

In no case was it possible to find any central drop for the velocity
dispersion. Of course, there is still a certain latitude in the fitting procedure,
but some features are certain: it is not possible to reproduce the
observations without a circumnuclear disc component, or with
an axisymmetric potential. Elliptical orbits are required, and the fits
give an order of magnitude of their importance. 
Also the required mass of the circumnuclear disc component is comparable,
and sometimes even greater, than the bulge mass.
The present models are simple first approximations, with bi-symmetry
imposed (there is no $m=1$ components, although in NGC~1808,
such an asymmetry is clearly observed); more realistic models 
constrained by further detailed kinematical data are needed to precise the 
dynamics of the double-bar galaxies. New models will also
help to examine the issue of the central mass concentrations,
for which we cannot, at the moment, give a lower limit.

\section{Discussion and conclusions}
   \label{sec:Conc}

We have presented the stellar kinematics of 4 galaxies hosting
an active nucleus, namely NGC~1097, NGC~1365, NGC~1808 and
NGC~5728, derived from \isaac/\vlt\ spectroscopy at 2.3~$\mu$m.

The essential results regarding the stellar kinematics of the
nuclear bars are the confirmation of the decoupling of the
nuclear component (with respect to the primary disc and bar), and 
the discovery of a central velocity dispersion drop in at least 
3 targets out of 4 (being unable to derive the central kinematics
for NGC~1365 due to the contribution of its Seyfert~1 nucleus). 
The observed central dispersion dips are not significantly affected when
optimal templates are used to derive the kinematics (Paper~II):
it is therefore a robust result.
We also observed a strong asymmetry in the stellar velocity 
profiles of NGC~1808, following the asymmetry in the photometry,
and suggesting the existence of an $m=1$ mode in the central
region of this galaxy.
The detailed discussion of the double-bar dynamics is postponed
to forthcoming papers, where it will be interpreted in terms 
of numerical simulations (through hydrodynamical N-body simulations 
and through determination of the orbital families with the 
Schwarzschild's method). In the following, we will discuss 
possible interpretations for the observed velocity dispersion drops.

The observation of a velocity dispersion drop at the centre of spiral
galaxies is rare. Such a drop has been observed in NGC~6503 by Bottema (\cite{Bott89}),
where the dispersion decreases within the central 12\arcsec.
The phenomenon could be more widespread at smaller radii, 
as it would be difficult to recognize it with limited
spatial resolution. In NGC~1097, the dip extends only 4\arcsec\ in radius, 
about $1\farcs5$ in NGC~1808, and $1\arcsec$ in the case of NGC~5728. 
The physical extent of the dispersion drop 
in NGC~1097 (radius of $\sim 4\arcsec \equiv 325$pc at 16.8~Mpc) is comparable 
to the one of NGC~6503 ($\sim 12\arcsec \equiv 350$pc at 6~Mpc),
but significantly larger than the ones in NGC~5728 
($\sim 1\arcsec \equiv 180$pc at 37~Mpc) and in NGC~1808
($\sim 1\farcs5 \equiv 80$pc at 10.9~Mpc).

Bottema (\cite{Bott93}) made a compilation of the velocity dispersion profiles
of a dozen spiral galaxies, and only NGC~6503 exhibits this drop. In general,
the dispersion profile is well fitted by an exponential law, decreasing
with a characteristic scale of twice the photometric scale-length for
the disc. When the bulge is significant, the fit is compatible with a constant
dispersion for the bulge. 
This exponential law for the disc is naturally explained for face-on
galaxies, i.e. for the vertical velocity dispersion profiles. Indeed,
it has been shown by van der Kruit \& Searle (\cite{Van81}, \cite{Van82}) that the
galactic discs have a constant scale-height with radius. Since
the surface density in the plane has an exponential distribution
(Freeman \cite{Free70}), the vertical equilibrium of a self-gravitating disc
implies that the dispersion varies as the square root of the surface
density, therefore in $e^{{-r}\over{2h}}$, where $h$ is the disc
radial scalelength. If the ratio between the radial and vertical
dispersions is maintained constant with radius, this will also
imply the same exponential behaviour for the in-plane dispersion.
Observations of the velocity profiles in inclined galaxies
seem to support this hypothesis of a constant ratio (Bottema~\cite{Bott93}). 

Also an interpretation in terms of disc stability and self-regulation
with the Toomre $Q$ parameter has been advanced (Bottema~\cite{Bott93}). Stars are heated
by gravitational instabilities like spirals and bars. When the $Q$ parameter
is too small, instabilities set in, until the velocity dispersion has increased
up to the threshold $Q$. The gaseous component allows a much richer
feedback regulation, since it can cool down through dissipation
and provoke recurrent instabilities. Young stars are formed out
of the gas with relatively low velocity dispersion. It is easy to see
how gravitational instabilities could lead to a constant $Q$ value all
over the stellar discs. 

Bottema \& Gerritsen (\cite{Bott97}) have re-examined the problem of the 
dispersion drop in the centre of NGC~6503, and find no 
intrinsic explanation. An hypothesis is to assume a very thin and cold disc in the centre,
but it is difficult to avoid heating of this disc through gravitational 
instabilities. They have undertaken N-body simulations to
check the stability of such a disc, and found only negative
results: no dispersion drop was ever observed in the simulations,
whatever the initial conditions. They conclude that the only 
solution is to assume the existence of an independent system in the
nucleus, a different population, that could have been recently accreted
from outside. The accretion must be quite recent.
Another explanation is the existence of two counter-rotating
bars, as suggested by Friedli (\cite{Fri96}). This hypothesis is not supported
by the observed kinematics in any of the three cases studied
in the present paper.

It could also be considered that fresh gas is radially falling inwards,
because of gravitational torques from a bar for instance, and that 
this gas is piling up in a thin
disc in the centre, then forming new stars with a low velocity
dispersion. There should have been then a recent starburst in the
centre of the galaxy. This scenario is likely for NGC~1808,
as we indeed detect a young stellar component in its centre
(Paper~II). The case of NGC~1097 may be
more difficult to assess. Kotilainen et al. (\cite{Kot00}) did find
some recent (6-7 Myr ago) star formation in the central region
of NGC~1097, but well distributed along its well-known (ILR) ring-like
structure. There is no evidence so far for a recent starburst inside the ring,
although we can not discard this hypothesis. 
New self-consistent N-body simulations including star formation
however support this scenario as an explanation for 
the observed central dispersion drop (Wozniak et al. 2001, in preparation).
We still need to understand how common this phenomenon is, among
galaxies with and without bars (single or double), and how it is linked to
the nuclear activity.

\begin{appendix}

\section{Formulae in the linear approximation}

In the absence of a non-axisymmetric perturbation, the orbits
can be computed in the epicyclic approximation, and the variables
 $\xi$, $\eta$ $<< r_0$ follow the evolution of an harmonic 
oscillator, with the epicyclic frequency $\kappa$. In the presence
of a bar perturbation, the equations of motions, in the 
reference frame rotating with the perturbation at $\Omega_p$ are
(Lindblad \& Lindblad \cite{Lin94}):
$$
\ddot{\xi}  - 2\Omega\dot{\eta}  - 4\Omega A\xi = 
{{d\Phi_2}\over{dr}} \cos{2\theta} = C  \cos{2\theta }
$$
$$
\ddot{\eta}  + 2\Omega\dot{\xi}  = -2{{\Phi_2}\over{r}} \sin{2\theta} = -D  \sin{2\theta }
$$
where $A$ is the Oort constant:
$$
A = -\frac{1}{2}r\frac{d\Omega}{dr},
$$
related to the epicyclic frequency $\kappa$ by:
$$
\kappa^2 = 4\Omega^2 -4\Omega A .
$$

The oscillator is now forced by an external perturbation at the 
imposed frequency $\omega = 2(\Omega-\Omega_p)$.

Taking for $\xi$ and $\eta$ a solution of the form:
$$
\xi = a \cos {(2\theta + 2\psi)} + c e^{-\lambda t} \cos{(\kappa^{'} t +\phi_0)}
$$
$$
\eta = b \sin {(2\theta + 2\psi)} + c^{'} e^{-\lambda t} \cos{(\kappa^{'} t +\phi_0)}
$$
with the same phase angle $\psi$ ($\xi$ and $\eta$ in quadrature),
and $\kappa^{'} $ the modified proper frequency of the damped
oscillations, it can be found that:

$$
a = {{\frac{d\Phi_2}{dr} + 4 \frac{\Omega}{\omega} \frac{\Phi_2}{r}}\over 
{\sqrt{(\kappa^2-\omega^2)^2+4\omega^2\lambda^2}}}
$$
and
$$
tan 2 \psi = -{{2 \omega\lambda}\over{\kappa^2-\omega^2}}
$$
with
$$
\omega^2 b = D -2\Omega\omega a
$$
The damped terms, exponentially decreasing with $\lambda t$, correspond
to the epicycles around the guiding centers, and are not considered here
(only through the velocity dispersion).

\end{appendix}
   
\begin{acknowledgements}
We wish to thank Jean-Gabriel Cuby and Claire Moutou
for their help and support during the \isaac\ observations. 
We also wish to thank the referee, Alan Morwood, for a detailed
and critical reading of the manuscript.
This work has been supported by the Swiss National Science Foundation.
\end{acknowledgements}

\end{document}